\documentclass[pra,aps,showpacs,twocolumn,floatfix,superscriptaddress]{revtex4-1}
\usepackage{graphicx}
\usepackage[ansinew]{inputenc}
\usepackage{array}
\usepackage{color}
\usepackage{amsmath}
\usepackage{amsxtra}
\usepackage{amstext}
\usepackage{amssymb}
\usepackage{latexsym}
\usepackage{dsfont}
\DeclareMathOperator{\sech}{sech}
\DeclareMathOperator{\csch}{csch}
\DeclareMathOperator{\sgn}{sgn}

\begin{document}
\title{Synthetic magnetism for photon fluids}
\author{N. Westerberg}
\email{nkw2@hw.ac.uk}
\affiliation{Institute of Photonics and Quantum Sciences, Heriot-Watt University, Edinburgh EH14 4AS, United Kingdom}
\author{C. Maitland}
\affiliation{Institute of Photonics and Quantum Sciences, Heriot-Watt University, Edinburgh EH14 4AS, United Kingdom}
\author{D. Faccio}
\affiliation{Institute of Photonics and Quantum Sciences, Heriot-Watt University, Edinburgh EH14 4AS, United Kingdom}
\affiliation{College of Optical Sciences, University of Arizona, Tucson, Arizona 85721, USA}
\author{K. Wilson}
\affiliation{Institute of Photonics and Quantum Sciences, Heriot-Watt University, Edinburgh EH14 4AS, United Kingdom}
\author{P. {\"O}hberg}
\affiliation{Institute of Photonics and Quantum Sciences, Heriot-Watt University, Edinburgh EH14 4AS, United Kingdom}
\author{E. M. Wright}
\affiliation{College of Optical Sciences, University of Arizona, Tucson, Arizona 85721, USA}
\affiliation{Institute of Photonics and Quantum Sciences, Heriot-Watt University, Edinburgh EH14 4AS, United Kingdom}


\begin{abstract}

We develop a theory of artificial gauge fields in photon fluids for the cases of both second-order and third-order optical nonlinearities. This applies to weak excitations in the presence of pump fields carrying orbital angular momentum and is thus a type of Bogoliubov theory. The resulting artificial gauge fields experienced by the weak excitations are an interesting generalization of previous cases and reflect the {\it PT}-symmetry properties of the underlying non-Hermitian Hamiltonian. We illustrate the observable consequences of the resulting synthetic magnetic fields for examples involving both second-order and third-order nonlinearities. 

\end{abstract}
\date{\today}
\maketitle

\section{Introduction}
The past two decades have seen a tremendous synergy between the fields of nonlinear optics and ultracold atomic gases. In particular, with the advent of atomic Bose-Einstein condensation \cite{Pethick} the area of nonlinear atom optics emerged \cite{Meystre} with clear similarities as well as some differences with traditional nonlinear optics \cite{Boyd}. This produced such fruitful concepts as matter-wave solitons, four-wave mixing, and second-harmonic generation in the form of coupled atom-molecule condensates \cite{KevFraGon}. More recently ideas from the matter-wave community have crossed into the nonlinear optics arena, in particular, the study of quantum fluids of light, or photon fluids \cite{Carusotto2013, Carusotto2014,Larre2015, Chiao1999,Weiss1996, Frisch1992}. The formal similarity between the Gross-Pitaevskii equation (GPE) for the macroscopic wave function of a two-dimensional Bose-Einstein condensate (BEC) and the paraxial wave equation for coherent propagation in a nonlinear medium is clear from the outset. Viewing the many photons forming the light field as a gas of Bose particles interacting via the medium nonlinearity reveals a new arena for the experimental and theoretical study of quantum many-body dynamics. At the forefront of studies of photon fluids are {\emph{polariton}} fluids, i.e. strongly coupled exciton-polaritons in semiconductor microcavities. These polariton fluids have been shown to exhibit superfluid behavior: Frictionless flow around an obstacle was demonstrated in Ref. \cite{Amo2009}, shedding of solitons was found in Ref. \cite{Amo2011}, and quantized vortices were studied in Refs. \cite{Sanvitto2011, Nardin2011}. These driven-dissipative systems with a {\emph{local}} nonlinearity are, however, not the only photon fluids. Room-temperature {\emph{nonlocal}} photon fluids in a propagating geometry have also displayed signatures of superfluid behavior in the dispersion relation. Specifically, the phonon-like linear dispersion of the long-wavelength collective modes was measured experimentally for coherent light propagation in a thermo-optical medium~\cite{VocRogMar15}, along with nucleation of quantized vortices in the flow past an extended physical obstacle \cite{VocWilMar16}. 

It is anticipated that as they are developed and refined, photon fluids will provide an alternative platform for some of the applications for quantum many-body systems that may operate at room temperature. In particular, proposals for quantum simulators based on cold atomic gases may be translated into the domain of photon fluids and offer considerably reduced technological expense. In these systems matter-wave or optical vortices would play the role of topologically protected quantum states \cite{LuJoaSol14,RagHal08,HalRag08,WanChoJoa08,WanChoJoa09,HafDemLuk11}. A key development in the matter-wave case has been artificial or synthetic gauge potentials, and associated synthetic magnetism, for neutral atoms \cite{DalGerJuz11} that are used to create and manipulate vortices within the quantum simulator and allow for the simulation of phenomena, such as the quantum Hall effect.

With the above discussion in mind the goal of the current paper is to develop the theory of artificial gauge potentials for photon fluids. A number of proposals and experiments for realizing artificial gauge fields for photons have arisen in the context of linear optics, including twisted \cite{Lon07,RecZeuPlo13} or strained \cite{RecZeuTun12,SchHal13} waveguide lattices, coupled-resonator optical waveguides \cite{HafDemLuk11,HafMitFan13,LiaCho13,MitFanFae14}, dynamic modulation of the parameters of photonic lattices \cite{FanYuFan12,Lon13,Lon14,TzuFanNus14}, driven dissipative lattices of polaritons \cite{UmuCar11,OzaCar14}, and optical cavities \cite{schine2015synthetic} with anamorphic optical elements \cite{Lon15}. Here our goal is to explore the idea of synthetic magnetism in the context of traditional nonlinear optics and in particular photon fluids. In the matter-wave case the simplest model for artificial gauge potentials is a two-level atom model in interaction with a laser field with a spatially tailored profile. To highlight the generic nature of our proposal we examine the analogous situation from nonlinear optics in two distinct cases: First we consider two orthogonal polarization basis states that are parametrically coupled in a medium displaying a second-order nonlinearity. Second, for the case of an isotropic medium displaying a third-order nonlinearity we consider the optical Bogoliubov-de Gennes equations for collective excitations around a strong solitary optical wave.  In Sec. \ref{MEqs} we provide the paraxial wave equations for these two cases and show that they may be expressed in a common quantum notation that will facilitate our discussion of the artificial gauge potentials for photon fluids by extension from the matter-wave case. Section \ref{AGP} develops the notion of artificial gauge potentials for photon fluids highlighting similarities and differences with the matter-wave case. Finally illustrative examples of the appearance of synthetic magnetic fields are given in Sec. \ref{Examples} for both second- and third-order nonlinearities, and our summary and conclusions are given in Sec. \ref{SumCon}.

\section{Model equations}\label{MEqs}

In this section we describe two nonlinear optical scenarios that lend themselves to a description in terms of artificial gauge potentials. Both cases involve the paraxial wave equations for fields propagating along the $z$ axis, the first involving a second-order nonlinearity, and the second a third-order nonlinearity. The resulting equations are put in a common quantum notation that facilitates introduction of the artificial gauge potentials by comparison to the matter-wave case.
\subsection{Second-order nonlinearity\label{chi2}}
In a second-order nonlinear medium the most general process is three-wave mixing in which fields at frequencies $\omega_{1,2}$ mix to produce a third field at $\omega_3=(\omega_1+\omega_2)$; this incorporates both sum and difference generation since the frequencies involved can have either sign. We consider the case with two fundamental fields, the signal and idler, of equal frequency $\omega_1=\omega_2=\omega$ but orthogonal polarization (type-II interaction) described by complex electric field envelopes ${\cal A}_{1,2}({\bf r})$ which are propagating along the $z$ axis. The third field is then at the second-harmonic $\omega_3=2\omega$ and is described by the complex electric-field envelope ${\cal A}_3({\bf r})$. Here it is assumed that all three fields are polarized along principal axes, and the $z$ axis is a principal axis so that there is no walk off. Then assuming that the signal (${\cal A}_{1}$) and idler (${\cal A}_{2}$) fields are weak in comparison to the strong pump field ${\cal A}_3$, the pump may then be assumed unchanged by the nonlinear interaction, leading to the following paraxial wave equations for the signal and idler field envelopes \cite{Boyd29},
\begin{eqnarray}
{\partial {\cal A}_1\over \partial z} &=& {i\over 2k_1} \nabla^2 {\cal A}_1 + \left ({2i\omega^2d_{\text{eff}}\over k_1c^2} \right ){\cal A}_2^* {\cal A}_3 e^{i\Delta k z} , \nonumber \\
{\partial {\cal A}_2\over \partial z} &=& {i\over 2k_2} \nabla^2 {\cal A}_2 + \left ({2i\omega^2d_{\text{eff}}\over k_2 c^2}\right ){\cal A}_1^* {\cal A}_3 e^{i\Delta k z} ,
\end{eqnarray}
where $k_{j}=n_{j}\omega_j/c$, $n_j$ being the refractive index for the given frequency and polarization state, $\nabla^2$ is the two-dimensional Laplacian on the $(x,y)$ plane, $\Delta k= (k_3-k_1-k_2)$ is the wave-vector mismatch along the $z$ axis, and $d_{\text{eff}}$ is the effective second-order nonlinear coefficient for the medium. Next we transform the fields according to
\begin{equation}
{\cal A}_{1,2}({\bf r}) = {a_{1,2}({\bf r})\over\sqrt{n_{1,2}}} e^{i\Delta kz/2} , \quad {\cal A}_3({\bf r}) = a_3({\bf r})e^{-i\phi({\bf r})} ,
\end{equation}
to obtain the propagation equations,
\begin{eqnarray}\label{aeqs}
i{\partial a_1\over \partial z} &=& -{1\over 2k_1} \nabla^2 a_1+{\Delta k\over 2}a_1 -\Gamma({\bf r}) e^{-i\phi}a_2^*, \nonumber \\
i{\partial a_2^*\over \partial z} &=& +{1\over 2k_2} \nabla^2 a_2^* -{\Delta k\over 2}a_2^*+\Gamma({\bf r}) e^{i\phi}a_1 ,
\end{eqnarray}
in which the wave-vector mismatch $\Delta k$ now appears as a term in the equations and the intensities are $n_j|{\cal A}_j|^2=|a_j|^2, j=1,2$. Here $\phi({\bf r})$ is the {\it phase angle} associated with the pump beam, and we assume that $a_3({\bf r})$ may vary with space but is real: The phase angle can therefore account for orbital angular momentum (OAM) of the pump field. Finally, we have defined the real function,
\begin{equation}
\Gamma({\bf r}) = {2\omega d_{\text{eff}}a_3({\bf r})\over\sqrt{n_1n_2} c}.
\end{equation}

We can convert the above Eqs. (\ref{aeqs}) for the signal and idler fields to quantum notation by multiplying though by ${\hbar v}$ with $v=c/\sqrt{n_1n_2}$ being the geometrical mean of the velocities of the fundamental fields, using $z=vt$ to convert from the propagation coordinate ($z$) to time, defining the effective photon mass $m$ via the relation $\hbar\omega=mv^2$, and using the spinor notation,
\begin{equation}
\left ( \begin{array}{c}
a_1 \\
a_2^*
\end{array}
\right ) \rightarrow
\left ( \begin{array}{c}
\psi_e \\
\psi_g
\end{array}
\right ) ,
\end{equation}
where the basis states $|e\rangle$ and $|g\rangle$ can be viewed as the excited and ground states of an effective two-level atom. The equations of motion then become in matrix form
\begin{equation}\label{Qchi2}
i\hbar {\partial \over \partial t} 
\left ( \begin{array}{c}
\psi_e \\
\psi_g
\end{array}
\right ) = 
\left ( \begin{array}{cc}
-{\hbar^2\kappa\over 2m}\nabla^2 & 0 \\
0 & + {\hbar^2\over 2m\kappa}\nabla^2
\end{array}
\right )
\left ( \begin{array}{c}
\psi_e \\
\psi_g
\end{array}
\right ) + U \left ( \begin{array}{c}
\psi_e \\
\psi_g
\end{array}
\right ) .
\end{equation}
Here the constant $\kappa=\sqrt{{n_2\over n_1}}$, and the {\it coupling operator} is given explicitly by
\begin{equation}\label{U}
U =\left ( \begin{array}{cc}
{\hbar v\Delta k\over 2} & -\hbar v\Gamma e^{-i\phi} \\
\hbar v\Gamma e^{i\phi} & - {\hbar v\Delta k\over 2} 
\end{array}
\right ) ,
\end{equation}
with eigenvalues $E_{1,2}=\pm \hbar\Omega/2$ and Rabi-frequency,
\begin{equation}\label{Omega}
\Omega = v\sqrt{\Delta k^2 - 4\Gamma^2}.
\end{equation}
Note that the eigenvalues are both real if $|\Delta k|>2|\Gamma|$, otherwise they are both imaginary. The condition $2|\Gamma|>|\Delta k|$ coincides with the appearance of parametric gain and loss for the system \cite{Boyd29}. For the case with real energy eigenvalues it is advantageous to write the interaction operator in the form
\begin{eqnarray}\label{Ureal}
U &=& {\hbar\Omega\over 2} \left ( \begin{array}{cc}
\sgn(\Delta k)\cosh(\theta) & -\sgn(\Gamma)\sinh(\theta) e^{-i\phi} \\
\sgn(\Gamma)\sinh(\theta) e^{i\phi} & -\sgn(\Delta k)\cosh(\theta) 
\end{array}
\right ) \nonumber \\ 
&=& {\hbar\Omega\over 2} \left ( \begin{array}{cc}
\zeta\cos\left (\xi \right ) & i\sin\left (\xi \right ) e^{-i\phi} \\
-i\sin\left (\xi \right ) e^{i\phi} & -\zeta\cos\left (\xi \right ) 
\end{array}
\right ) ,
\end{eqnarray}
with {\it complex mixing angle} $\xi=i\theta$ and
\begin{equation}\label{thetaReal}
\tanh(\theta) = \left |{2\Gamma\over\Delta k} \right | .
\end{equation}
Here $\zeta=\sgn(\Delta k)$, and the factor $\sgn(\Gamma)$ may be subsumed into the phase angle via the replacement $\phi\rightarrow \phi+\pi$ if $\Gamma<0$. The occurrence of complex mixing angles will be discussed further in Sec. \ref{AGP}.

For the case of $2|\Gamma|>|\Delta k|$ with imaginary eigenvalues it will prove useful to express the interaction operator in the form
\begin{eqnarray}\label{Uimag}
U &=& {\hbar\Omega\over 2} \left ( \begin{array}{cc}
-i\zeta\sinh(\theta) & i\sgn(\Gamma)\cosh(\theta) e^{-i\phi} \\
-i\sgn(\Gamma)\cosh(\theta) e^{i\phi} & i\zeta\sinh(\theta) 
\end{array}
\right ) \nonumber \\
&=& {\hbar\Omega\over 2} \left ( \begin{array}{cc}
\zeta\cos\left (\xi \right ) & i\sin\left (\xi \right ) e^{-i\phi} \\
-i\sin\left (\xi \right ) e^{i\phi} & -\zeta\cos\left (\xi \right ) 
\end{array}
\right ) ,
\end{eqnarray}
with $\xi=(i\theta+\pi/2)$ and $\theta$ determined by
\begin{equation}\label{thetaImag}
\tanh(\theta) = \left | {\Delta k\over 2\Gamma} \right | .
\end{equation}
The benefit of this notation is that the interaction operator assumes the same functional form for the cases of both real and imaginary eigenvalues, the differences between the two cases residing in the expressions for the mixing angles $\xi$ and $\theta$.

Equation (\ref{Qchi2}) has a formal similarity to those previously considered for a matter-wave system composed of two-level atoms \cite{DalGerJuz11}. In particular, for the case of both real and imaginary eigenvalues the factor $\Omega\cos(\xi)$ plays the role of the atom-laser detuning, and $\Omega\sin(\xi)$ is the magnitude of the atom-laser coupling. In comparison to the matter-wave case, the second-harmonic pump field ${\cal A}_3$ corresponds to a molecular condensate whereas the signal and idler, which are cross coupled via interactions with the molecular condensate, correspond to atomic condensates in different hyperfine ground states \cite{donely2002}. In this sense our model for parametric interaction of the signal and idler fields can be viewed as an example of a two-component or spinor photon fluid. For simplicity in presentation we have not included the usual $s$-wave interactions that would appear for each individual condensate, or Kerr nonlinearity in the nonlinear optics case as that is the topic of the next subsection. 
\subsection{Third-order nonlinearity}
In this case we consider propagation of a linearly polarized and monochromatic field of frequency $\omega$ in an isotropic nonlinear medium with spatially varying linear refractive-index $n({\bf r})=n_0+\Delta n({\bf r})$ with $n_0$ as the background refractive index and $\Delta n$ describes the spatial profile of small variations in the index. Then within the paraxial approximation the equation for the scalar slowly varying electric-field envelope is \cite{Boyd71}
\begin{equation}\label{NLS}
{\partial {\cal A}\over \partial z} ={i\over 2k_0} \nabla^2 {\cal A}+i\left ({\omega\over c} \right )\Delta n{\cal A}+i\left ( {3\chi^{(3)}\omega\over 2n_0 c} \right )|{\cal A}|^2{\cal A} ,
\end{equation}
with $\chi^{(3)}$ as the third-order nonlinear susceptibility which plays the role of the scattering length that gauges the strength of $s$-wave interactions in the BEC case. For a cylindrically symmetric refractive-index profile $\Delta n(\rho)$ this equation permits solitary wave solutions carrying OAM of the form ${\cal A}({\bf r}) = {\cal A}_0(\rho)e^{i\beta z+i\ell \varphi}$ in cylindrical coordinates ${\bf r}=(\rho,\varphi,z)$ such that
\begin{equation}
\beta {\cal A}_0 = {1\over 2k_0} \left ({d^2\over d\rho^2} + {1\over \rho}{d\over d\rho} -{\ell^2\over \rho^2} \right ){\cal A}_0 +\left ({\omega\over c} \right )\Delta n {\cal A}_0 + \eta {\cal A}_0^3 ,
\end{equation}
with ${\cal A}_0(\rho)$ real, $\rho=\sqrt{x^2+y^2}$, and $\eta=\left ( {3\chi^{(3)}\omega\over 2n_0 c} \right )$. Here we are interested in small fluctuations around the solitary wave solution, akin to the collective excitations appearing in the Bogoliubov theory for matter waves \cite{Pethick8} and set
\begin{equation}
{\cal A}({\bf r}) = e^{i\beta z}[{\cal A}_0(\rho)e^{-i\phi({\bf r})/2} + a_+({\bf r})+a_-({\bf r})] ,
\end{equation}
where $\phi({\bf r})=-2\ell\varphi$. The fluctuations $a_\pm({\bf r})$ are taken to be distinguishable via their transverse mode structure.  For our particular application in which the phase angle $\phi({\bf r})$ describes a strong solitary wave carrying OAM, $a_\pm$'s correspond to orthogonal states with OAM related by $\ell_++\ell_-=2\ell$.  Then linearizing with respect to the fluctuations $a_\pm({\bf r})$ yields the {\it optical Bogoliubov-de Gennes equations} for the photon fluid,
\begin{eqnarray}\label{aeq2}
i{\partial a_+\over \partial z} &=& -{1\over 2k_0} \nabla^2 a_+-\left ({\omega\over c} \right )\Delta n a_++ {\Delta k\over 2} a_+-\Gamma e^{-i\phi} a_-^*, \nonumber \\
i{\partial a_-^*\over \partial z} &=& {1\over 2k_0} \nabla^2 a_-^*+ \left ({\omega\over c} \right )\Delta n a_-^*- {\Delta k\over 2} a_-^*+ \Gamma e^{i\phi} a_+,
\end{eqnarray}
where we have defined
\begin{equation}
{\Delta k\over 2} = \beta +2 \eta {\cal A}_0^2 , \quad \Gamma = \eta {\cal A}_0^2 .
\end{equation}

At this point the similarity between Eqs. (\ref{aeqs}) for the parametric interaction and Eqs. (\ref{aeq2}) for the present case is obvious, and we may convert to the quantum notation as before using $v=c/n_0$, defining the potential,
\begin{equation}
V({\bf r}) = -\hbar v \left ({\omega\over c} \right )\Delta n({\bf r}) ,
\end{equation}
along with the spinor notation,
\begin{equation}\label{spinor}
\left ( \begin{array}{c}
a_+ \\
a_-^*
\end{array}
\right ) \rightarrow
\left ( \begin{array}{c}
\psi_e \\
\psi_g
\end{array}
\right ) .
\end{equation}
The equations of motion for this case can then be written in the common quantum notation as
\begin{eqnarray}\label{Qchi3}
i\hbar {\partial \over \partial t} 
\left ( \begin{array}{c}
\psi_e \\
\psi_g
\end{array}
\right ) &=&
\left ( \begin{array}{cc}
{{\bf P}^2 \over 2m}+V & 0 \\
0 & - {{\bf P}^2 \over 2m} - V
\end{array}
\right )
\left ( \begin{array}{c}
\psi_e \\
\psi_g
\end{array}
\right ) + U \left ( \begin{array}{c}
\psi_e \\
\psi_g
\end{array}
\right ) \nonumber \\
&=&
\sigma_3 \left ( {{\bf P}^2 \over 2m}+V \right )
\left ( \begin{array}{c}
\psi_e \\
\psi_g
\end{array}
\right ) + U \left ( \begin{array}{c}
\psi_e \\
\psi_g
\end{array}
\right ) ,
\end{eqnarray}
with ${\bf P}=-i\hbar\boldsymbol{\nabla}$, $\sigma_3$ as the Pauli spin-matrix, and all other parameters defined as before in Eqs. (\ref{U})-(\ref{thetaImag}). In our quantum notation the only differences between Eqs. (\ref{Qchi2}) and (\ref{Qchi3}) for the two nonlinear optics examples is the factor $\kappa=\sqrt{n_2/n_1}$ that appears in the parametric model due to the different phase velocities of the signal and idler and the addition of the refractive-index profile leading to the potential $V$. However, for our purposes $\sqrt{n_2/n_1}\sim 1$, otherwise the nonlinear interaction would be too highly phase mismatched. We therefore adopt Eq. (\ref{Qchi3}) as our generic model in our discussion of artificial gauge potentials for photon fluids in the next section.

For the above example our starting nonlinear paraxial wave equation (\ref{NLS}) is analogous to the GPE for a two-dimensional BEC, and the linearized theory for the fluctuations $a_\pm$ corresponds to the Bogoliubov theory for the collective excitations \cite{Pethick8}.  To conclude we remark that it is possible to combine both second- and third-order nonlinear interactions together, so that the resulting photon fluid incorporates Kerr-type interactions of the signal and idler fields as well as a cross interaction mediated by the second-harmonic field (or in BEC terms it includes $s$-wave interactions in addition to atom-molecule coupling). However, for simplicity in notation we have chosen to illustrate these separately. 
\section{Artificial gauge potentials}\label{AGP}
In this section we describe the introduction of artificial gauge potentials using Eq. (\ref{Qchi3}) as our starting point. Our treatment follows that of Ref. \cite{DalGerJuz11} for matter waves, and we stress where key differences arise.
\subsection{Dressed states}
The main concept needed for introducing artificial gauge potentials is the use of the dressed states of the interaction operator $U$, obtained as the eigenstates of this operator. This leads to a key difference with respect to the matter-wave case: In the present setting $U$ is not a Hermitian operator. On the other hand, the associated dressed state eigenvalues are $E_{1,2}=\pm {\hbar\Omega\over 2}$, which are either both real for $|\Delta k|\ge 2|\Gamma|$ or both imaginary for $|\Delta k|< 2|\Gamma|$, see Eq. (\ref{Omega}). This is reminiscent of the class of Hamiltonians that are non-Hermitian but do display parity-time ($PT$) symmetry and can yield real eigenvalues \cite{BenBot98,Guo09}. More specifically, following Bender {\it et al.} \cite{BenBerMan02} the combined action of the parity operator $P$, which interchanges the basis states $|e\rangle\leftrightarrow|g\rangle$, and the time-reversal operator $T$, which takes the complex conjugate, on the interaction operator yields $[PTU]_{\mu\nu}=U_{\nu\mu}^*$, with $\mu,\nu=\pm 1$ and the identifications $+1\equiv e$, $-1\equiv g$. Then for the case with $|\Delta k|>2|\Gamma|$ with real eigenvalues we find $PTU=-U$, the real eigenvalues giving rise to a {\it phase-conjugate} coupling between the basis states with concomitant oscillatory dynamics. In this case the interaction operator displays {\it anti-$PT$ symmetry} as recently revealed for parametric interactions in nonlinear optics \cite{AntSolSuk15}. In contrast, for the case $|\Delta k|<2|\Gamma|$ with imaginary eigenvalues $PTU=U$, and the system displays $PT$ symmetry, or broken anti-$PT$ symmetry. In this case the imaginary eigenvalues produce two-mode {\it parametric amplification} and loss.

The non-Hermitian nature of the interaction operator requires that we look at the dressed states provided by the right and left eigenstates of the interaction operator $|u_i\rangle$ and $\langle v_j|$, which are the dual of each other, and obey
\begin{equation}
U|u_j \rangle=E_j |u_j \rangle, \quad \langle v_j | U=E_j \langle v_j | .
\label{rabi}
\end{equation}
For concreteness we consider the case with $\zeta=\sgn(\Delta k)=1$, then for the eigenvalue $E_1=\hbar\Omega/2$ the right and left eigenstates are
\begin{eqnarray}
|u_1\rangle &=&\left ( \begin{array}{c}
\cos(\xi/2) \\
-i\sin(\xi /2)e^{i\phi}
\end{array}
\right ) , \nonumber \\
\langle v_1|&=& (
\cos(\xi/2),~ 
i\sin(\xi/2)e^{-i\phi}) ,
\end{eqnarray}
whereas for $E_2=-\hbar\Omega/2$,
\begin{eqnarray}
|u_2\rangle &=&\left ( \begin{array}{c}
\sin(\xi /2) \\
i\cos(\xi /2)e^{i\phi}
\end{array}
\right ) , \nonumber \\
\langle v_2|&=& 
(\sin(\xi /2),~ 
-i\cos(\xi /2)e^{-i\phi} ) .
\end{eqnarray}
For the case with $\zeta=\sgn(\Delta k)=-1$ one simply interchanges $E_1\leftrightarrow E_2$ and replaces $\zeta\rightarrow -\zeta$ in the eigenstates, see $U$ in Eq. (\ref{Uimag}). It is easily checked that these dressed states are biorthogonal in the sense $\langle v_i|u_j\rangle =\delta_{ij}$, and satisfy the completeness relation,
\begin{equation}\label{CR}
I = |u_1\rangle\langle v_1| + |u_2\rangle\langle v_2| ,
\end{equation}
with $I$ as the unit $(2\times 2)$ matrix. We will employ these properties in the following analysis.
\subsection{Artificial gauge potentials}
To start we expand the state vector generally as
\begin{equation}
|\psi \rangle = \left ( \begin{array}{c}
\psi_e \\
\psi_g
\end{array}
\right ) = \sum_{j=1}^2 \psi_j ({\bf r},t) |u_j\rangle ,
\label{stat}
\end{equation}
which yields
\begin{eqnarray}
{\bf P}|\Psi \rangle &=& -i\hbar \sum_j [\boldsymbol{\nabla} \psi_ j |u_j\rangle + \psi_j |\boldsymbol{\nabla} u_j\rangle ] \nonumber \\
&=& -i\hbar \sum_j \underbrace{\left [ \sum_l |u_l\rangle \langle v_l| \right ]} [\boldsymbol{\nabla} \psi_ j |u_j\rangle + \psi_j |\boldsymbol{\nabla} u_j\rangle ] \nonumber \\
&=& \sum_{j,l=1}^2 \left [ \right ( \delta_{jl}{\bf P}-{\bf A}_{jl}\left)\psi_l\right ]|u_j\rangle ,
\end{eqnarray}
where ${\bf A}_{jl}=i\hbar\langle v_j|\boldsymbol{\nabla} u_l\rangle$, and the completeness relation (\ref{CR}) was used in the underbraced term in the second line. A second application of this procedure then yields
\begin{equation}
\mathbf{P}^2|\Psi \rangle = \sum_{j,k,l=1}^2 \left [ (\delta_{jk} {\bf P} - {\bf A}_{jk})\cdot (\delta_{kl} {\bf P} - {\bf A}_{kl})\psi_l \right ] |u_j \rangle .
\label{psqr}
\end{equation}
The next step is to apply these results within the adiabatic approximation according to which the state vector remains dominantly in one of the dressed states, say $|u_1\rangle$, during its evolution. Then assuming that $\psi_2\sim 0$ remains negligible in the above equations, we project the Schr{\"o}dinger equation $i\hbar|\dot\Psi \rangle=H|\Psi\rangle$ in Eq. (\ref{Qchi3}) onto the dressed state $|u_1\rangle$ using the projector $\langle v_1|$ to obtain
\begin{eqnarray}\label{psi11}
i\hbar{\partial\psi_1\over\partial t} &=& \cos(\xi)\left [{({\bf P}-{\bf A}_{11})^2\over 2m} + V + W_1 \right ]\psi_1 + {\hbar\Omega\over 2}\psi_1 \nonumber \\
&-& {\sin(\xi)\over 2m} \left [ {\bf A}_{21} \cdot {\bf P} +{\bf P}\cdot {\bf A}_{21} -
{\bf A}_{21}\cdot\left( {\bf A}_{11}+{\bf A}_{22} \right ) \right ]\psi_1. \nonumber \\
\end{eqnarray}
Here the factor $\cos(\xi)$ multiplying the square brackets is a result of the projection $\langle v_1|\sigma_3|u_1\rangle$, whereas the factor $\sin(\xi)$ arises from the projection $\langle v_1|\sigma_3|u_2\rangle$. Furthermore the various artificial vector potentials are evaluated as
\begin{eqnarray}
{\bf A}_{11} &=& +{\hbar\over 2}(\cos(\xi)-1)\boldsymbol{\nabla}\phi , \nonumber \\
{\bf A}_{22} &=& -{\hbar\over 2}(\cos(\xi)+1)\boldsymbol{\nabla}\phi , \nonumber \\
{\bf A}_{12} &=& +{i\hbar\over 2}\boldsymbol{\nabla}\xi + {\hbar\over 2}\sin(\xi)\boldsymbol{\nabla}\phi, \nonumber \\
{\bf A}_{21} &=& -{i\hbar\over 2}\boldsymbol{\nabla}\xi + {\hbar\over 2}\sin(\xi)\boldsymbol{\nabla}\phi,
\end{eqnarray}
and the geometric scalar potential is
\begin{eqnarray}
W_1 &=& {1\over 2m} {\bf A}_{21}\cdot{\bf A}_{12} \nonumber \\
&=& {\hbar^2\over 8m} \left [ (\boldsymbol{\nabla}\xi)^2 + \sin^2(\xi)(\boldsymbol{\nabla}\phi)^2\right ] ,
\end{eqnarray}
The top line of Eq. (\ref{psi11}) is quite similar to the matter-wave result \cite{DalGerJuz11}, but there is a term $\cos(\xi)$ multiplying the Hamiltonian.

To proceed Eq. (\ref{psi11}) can be rearranged including the terms in the bottom line into the form
\begin{equation}
i\hbar{\partial\psi_1\over\partial t} = \cos(\xi)\left [{({\bf P}-{\bf A})^2\over 2m} + V + W\right ]\psi_1 + {\hbar\Omega\over 2}\psi_1 ,
\label{eom1}
\end{equation}
where the vector potential is given by
\begin{equation}\label{A}
{\bf A} ={\bf A}_{11} +\tan(\xi) {\bf A}_{21} ,
\end{equation}
and the geometric scalar potential is
\begin{eqnarray}\label{W}
W &=& W_1 + {1\over 2m}\left [ \tan(\xi){\bf A}_{21}\cdot ({\bf A}_{22}-{\bf A}_{11}) -\tan^2(\xi){\bf A}_{21}^2 \right ] \nonumber \\
&=& {\hbar^2\over 8m} \sec^2(\xi)\left [ (\boldsymbol{\nabla}\xi)^2 - \sin^2(\xi)(\boldsymbol{\nabla}\phi)^2 \right ] ,
\end{eqnarray}
Here we have assumed that $\boldsymbol{\nabla}\xi=i\boldsymbol{\nabla}\theta$ and $\boldsymbol{\nabla}\phi$ are orthogonal based on the fact that we are mainly interested in solutions with cylindrically symmetric intensity profiles that carry OAM.

If the same calculation is repeated for the case that the quantum system stays adiabatically in the dressed state $|u_2\rangle$ we obtain
\begin{equation}
i\hbar{\partial\psi_2\over\partial t} = -\cos(\xi)\left [{({\bf P}-{\bf A})^2\over 2m} + V + W\right ]\psi_2 - {\hbar\Omega\over 2}\psi_2 ,
\label{eom2}
\end{equation}
where the vector potential is given by
\begin{equation}
{\bf A} ={\bf A}_{22} +\tan(\xi) {\bf A}_{12} ,
\end{equation}
and the geometric scalar potential is the same as before. The factor $-\cos(\xi)$ multiplying the Hamiltonian in Eq. (\ref{eom2}) results from the projection $\langle v_2|\sigma_3|u_2\rangle$, and the negative sign follows physically from the fact that the dressed state $|u_2\rangle$ is dominated by the negative energy state in the state vector for small $\xi$.
\subsection{Validity of the adiabatic approximation}

Equations (\ref{eom1}) and (\ref{eom2}) are strictly speaking only valid when the population is assumed to remain mostly in state $|u_1\rangle$ or $|u_2\rangle$, respectively. This assumption holds as long as the coupling between the two states is weak. By explicitly calculating the matrix elements $\langle v_1|{\bf P}^2|\Psi\rangle$ from Eq. (\ref{psqr}) and identifying the term proportional to $\psi_2(x)$ we obtain the coupling strength between the coefficients $\psi_1$ and $\psi_2$ in the expansion $|\psi \rangle = \sum_{j=1}^2 \psi_j ({\bf r},t) |u_j\rangle$. From the resulting two-component Schr{\"o}dinger equation and Eqs. (\ref{eom1}) and (\ref{eom2}), we see that the dominating energy scale is the Rabi frequency $|\Omega|=|E|/\hbar$ in Eq. (\ref{rabi}), hence if $|{\bf P}\cdot {\bf A}_{12}|/m\ll\hbar|\Omega|$ and $|{\bf A}_{11}\cdot {\bf A}_{12}|/2m\ll \hbar|\Omega|$ are fulfilled, we can safely assume the adiabatic approximation holds.
\subsection{Discussion}
Before proceeding to specific examples of the artificial gauge potentials we will make some general observations. First, the gauge potentials do not depend on the sign of $\Gamma$ since only $\boldsymbol{\nabla}\phi$ appears, which is insensitive to the additional factor of $\pi$ used when $\Gamma<0$. Second, the vector potential in Eq. (\ref{A}) may be expressed as
\begin{equation}
{\bf A} = {\hbar\over 2}(\cos(\xi)-1)\boldsymbol{\nabla}\phi + {\hbar\over 2} \tan(\xi)\left ( -i\boldsymbol{\nabla}\xi + \sin(\xi)\boldsymbol{\nabla}\phi \right ) .\end{equation}
For the phase-conjugate case $|\Delta k|>2|\Gamma|$ with real energy eigenvalues (anti-$PT$ symmetry) we have $\xi=i\theta$, giving
\begin{equation}\label{AaPT}
{\bf A} = -{\hbar\over 2}(1-\sech(\theta))\boldsymbol{\nabla}\phi - {i\hbar\over 2} \tanh(\theta)\boldsymbol{\nabla}\theta ,
\end{equation}
with $\tanh(\theta)$ given by Eq. (\ref{thetaReal}). Then for $\boldsymbol{\nabla}\theta=0$ we find a real vector potential ${\bf A}$ but with zero accompanying artificial magnetic field ${\bf B}=\boldsymbol{\nabla}\times{\bf A}=0$. This will yield a real valued flux tube for phase angles $\phi({\bf r})$ carrying OAM in the phase-conjugate regime. Note that the associated flux can take on both integer and non-integer values. In contrast, for the parametric amplification case of $|\Delta k|<2|\Gamma|$ with imaginary energy eigenvalues ($PT$ symmetry) we have $\xi=i\theta+\pi/2$, leading to
\begin{equation}\label{APT}
{\bf A} = -{\hbar\over 2}(1+i\csch(\theta))\boldsymbol{\nabla}\phi - {i\hbar\over 2}{\boldsymbol{\nabla}\theta\over\tanh(\theta)} ,
\end{equation}
with $\tanh(\theta)$ given by Eq. (\ref{thetaImag}). In this case if $\boldsymbol{\nabla}\theta=0$ we find a {\it complex vector potential} ${\bf A}$ but still with zero accompanying artificial magnetic field. What is new is that this will yield a {\it complex valued flux tube} for phase angles $\phi({\bf r})$ carrying OAM in the regime of parametric amplification. The real part of this complex flux tube corresponds to an integer flux.

This is an interesting situation where the artificial flux tube may be real or complex, and is clearly related to the non-Hermitian character of the interaction operator $U$. From a quantum-mechanical point of view the artificial flux tube arises from an Aharonov-Bohm or Berry phase acquired as a particle executes a closed contour ${\cal C}$ enclosing the phase singularity,
\begin{equation}
\gamma({\cal C}) = {1\over\hbar}\oint_{\cal C} {\bf A}\cdot d{\bf r} .
\end{equation}
The Berry phase has a geometric origin and does not depend on the choice of contour. For the phase-conjugate case with real eigenvalues this Berry phase is real leading to a real-valued flux tube \cite{Berry}: In this case the system exhibits anti-$PT$ symmetry and acts like a Hermitian system for all intents and purposes. In contrast, for the parametric amplification case, the full non-Hermitian character of the problem is apparent as reflected in the imaginary eigenvalues, and this leads to a complex Berry phase and concomitant complex flux tube \cite{GarWri88}: In this case the anti-$PT$ symmetry is broken. The complex flux tube therefore encodes contributions to the amplification or loss that are purely geometric in nature, and are in addition to the dynamical amplification and loss included in the imaginary eigenvalues. It has previously been found for a two-level system model with loss that the complex Berry phase can be phrased in terms of complex solid angles, or mixing angles in our case, and that is why using complex angles has proven so fruitful in this derivation \cite{GarWri88}. 
\section{Synthetic magnetism}\label{Examples}

So far we have seen that second-order ($\chi^{(2)}$ case) and third-order ($\chi^{(3)}$ case) nonlinearities can give rise to synthetic gauge potentials in the presence of optical fields carrying OAM. That a synthetic vector potential $\bf{A}$ can appear for the $\chi^{(3)}$ case is not new, and previous work involving scattering of a dark soliton from an optical vortex uncovered the nonlinear Aharonov-Bohm effect \cite{KivNepTik00,NesNepKiv01}. A similar effect was also seen for scattering of acoustic waves from arrays of matter-wave vortices \cite{CouWei15}, a situation analogous to the $\chi^{(3)}$ case. In these cases, however, there is no associated magnetic field, that is, $\bf{B}=\boldsymbol{\nabla}\times\bf{A}=0$. We therefore consider illustrative examples below for which synthetic magnetic fields arise and point to observable consequences of their presence. 

\subsection{Second-order nonlinearity}

\begin{figure}
\includegraphics[width=8cm]{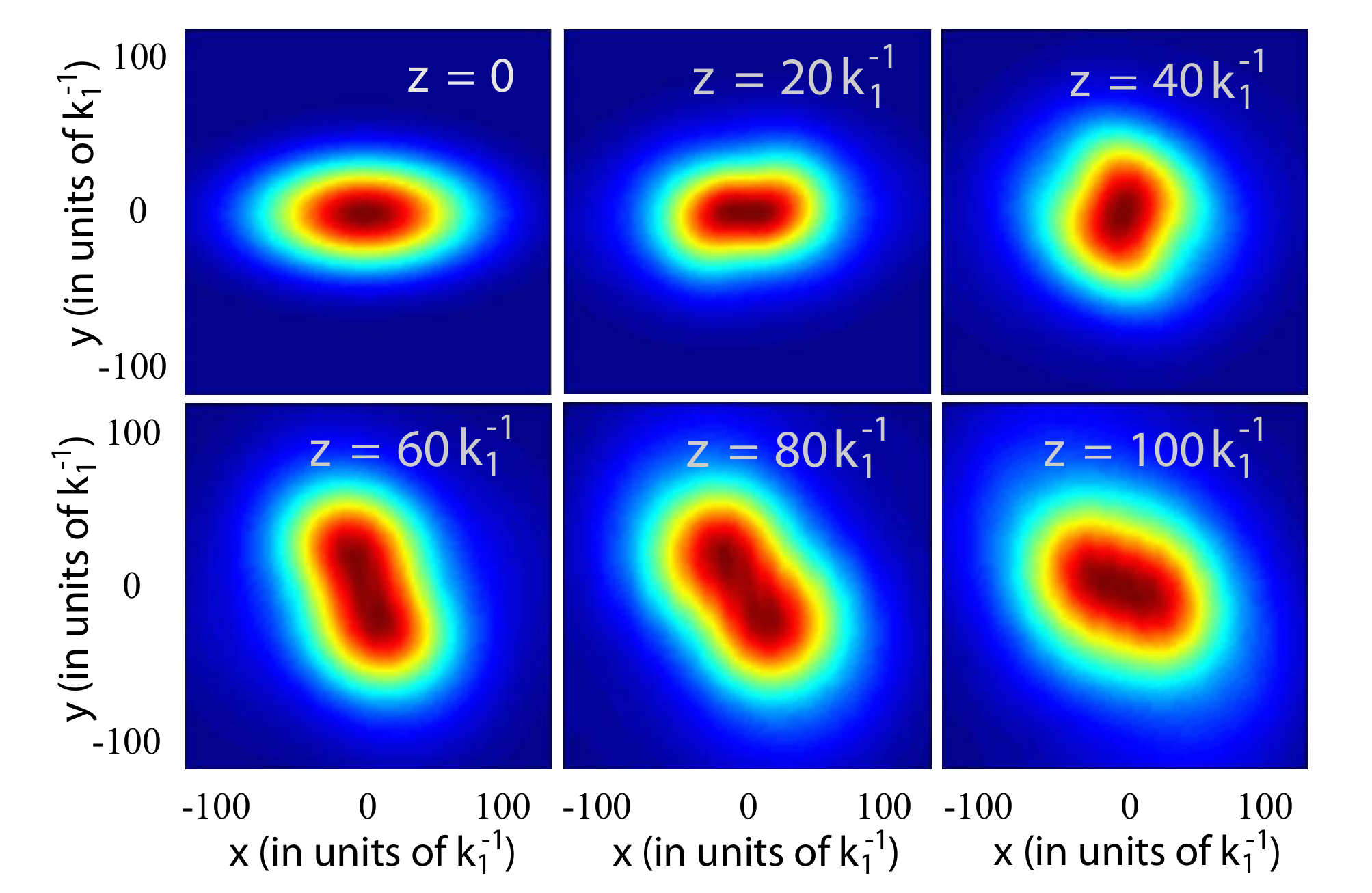}
\caption{Snapshots of the transverse intensity profile of the light beam as a function of $z$. The input beam at $z=0$ is an elliptic Gaussian, with the widths $\sigma_X = 40k_1^{-1}$, $\sigma_Y = 20k_1^{-1}$, which will rotate due to the nearly uniform artificial magnetic field in the center of the beam. The dynamics is described by a $\chi^{(2)}$ nonlinearity where the pump beam is a Laguerre-Gaussian beam with winding number $\ell=1$ and amplitude $\Psi(r)=\sqrt{5} (r/\sigma)\exp[-(r/\sigma)^2]$ where $\sigma=50k_1^{-1}$. The resulting intensity shown in the figure is a superposition of the signal and idler components, which forms the dressed state.}
\label{rot1}
\end{figure}

In order to illustrate the effect an artificial magnetic field has on the photon fluid, we will study the rotation induced by the synthetic magnetic field. A uniform magnetic field corresponds to an effective rotation of the coordinate system. We should therefore expect to see any shape of the photon fluid that breaks the rotational symmetry to rotate on the transverse plane with respect to the propagation axis, which is in the $z$ direction. To emulate an effective uniform magnetic field we will first consider the $\chi^{(2)}$ case in Section \ref{chi2}. We assume we are in the limit of real Rabi frequencies with $|\Delta k|>2|\Gamma|$. It is instructive to first consider the limit $|\Delta k|\gg 2|\Gamma|$ where we keep terms up to quadratic order in $2\Gamma/\Delta k$. By choosing a pump beam with an OAM $\hbar \ell$ one obtains a {\it complex} vector potential of the form 
\begin{equation}
{\bf A}=-\frac{\hbar}{4}\bigg(\frac{2\Gamma}{\Delta k}\bigg)^2\ell\frac{1}{r}\hat e_\theta+i\frac{\hbar}{2}\frac{2\Gamma}{\Delta k}\boldsymbol{\nabla}\left(\frac{2\Gamma}{\Delta k}\right)\label{vecpot2}
\end{equation}
where 
\begin{equation}
\frac{1}{2}\bigg(\frac{2\Gamma}{\Delta k}\bigg)=\sqrt{\alpha}r,
\end{equation}
with an effective magnetic field in the $z$ direction,
\begin{equation}
{\bf B}=\boldsymbol{\nabla}\times {\bf A}=-2\hbar\alpha\ell \hat e_z.\label{magfield}
\end{equation}
The parameter $\alpha$ captures the strength of the magnetic field. It has the unit of inverse length squared where the length in question is the characteristic length over which the pump beam amplitude is linear in $r$.
The second term on the right hand side of Eq. (\ref{vecpot2}) is imaginary but in this particular example purely radial in direction where $\Gamma$ is also a function of $r$ only. This term can therefore be gauged away. Near the center of the pump beam the intensity will be linear in $r$ for $\ell=1$. The resulting synthetic magnetic field in Eq. (\ref{magfield}) will then be uniform and cause a rotation of the photon fluid provided that the pump beam width is much larger than the dressed states formed by the signal and the idler beams. Figure \ref{rot1} shows how an elliptical input beam consisting of a particular combination of the signal and the idler beam, i.e., a dressed state, will experience a rotation while propagating along the $z$ direction. The figure shows a series of snapshots at different $z$ values. There are two independent contributing factors to the dynamics: slow expansion and an overall rotation of the whole system. The former is simply a diffraction effect due to the fact that the more tightly confined direction expands more rapidly, whereas the latter would indicate that there is an effective magnetic field acting on the photon fluid. It is instructive to note that changing the sign of the OAM, i.e., $\ell \rightarrow -\ell$, will simply flip the sign of the synthetic magnetic field. An elliptical beam, such as the input transverse intensity profile in Fig.~\ref{rot1}, would rotate clockwise instead of counterclockwise (data not shown). In this particular example we have explicitly created an almost uniform synthetic magnetic field, but we are not restricted to this choice. For instance a higher-order Laguerre-Gaussian beam with $\ell>1$ will give a radially dependent magnetic field in the $z$ direction. Alternatively we can take into account the full envelope of the pump beam such that the full ring-shaped amplitude is relevant. This will give a uniform magnetic field in the central region of the pump beam which will decay to zero once we are beyond the linear in $r$ dependence of the pump beam.

\subsection{Third-order nonlinearity}

\begin{figure}
\includegraphics[width=8cm]{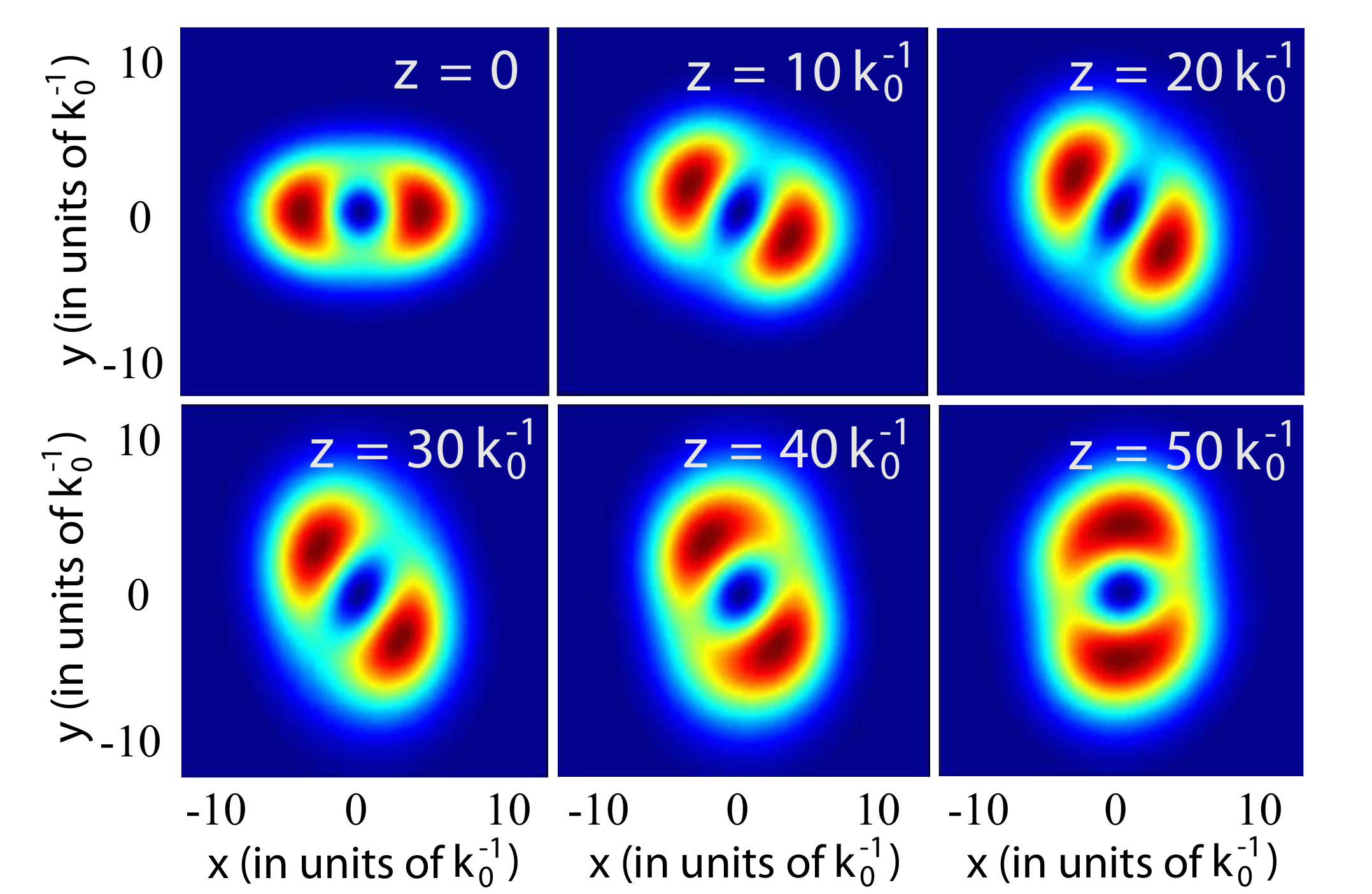}
\caption{Snapshots of the transverse intensity profile of the light beam as a function of $z$. The dynamics is governed by a single component nonlinear Schr{\"o}dinger equation with a $\chi^{(3)}$ nonlinearity. The input beam at $z=0$ is an elliptic Laguerre-Gaussian beam with winding number of $\ell=1$, $\sigma_{X} = 30k_0^{-1}$, $\sigma_Y = 20k_0^{-1}$ and peak intensity-dependent refractive index of $\delta n\simeq 0.0087$. The underlying optical vortex causes a rotation of the excitation axis. In other words, the excitation exhibits cyclotron motion. This can be interpreted as having an artificial magnetic field acting on the photon fluid.}
\label{chi3dyn}
\end{figure}

In the $\chi^{(3)}$ case we consider a similar situation as above. Here we have a strong pump beam subject to a nonlinearity proportional to the intensity. The dynamics is given by Eq. (\ref{NLS}). The corresponding optical Bogoliubov-de Gennes equations have the same structure as the $\chi^{(2)}$ case for the signal and idler beams. The difference is that in the $\chi^{(3)}$ case the excitations of the photon fluid will be subject to a synthetic magnetic field. It is worth noting at this point that a similar technique was used by Chevy \textit{et al.} \cite{chevy_2000} where they measured the angular momentum of a Bose-Einstein condensate by looking at the rotation of the quadrupole excitation axis and by doing so were able to deduce that there was a quantized vortex. In their experiment a vortex was created in the BEC. Subsequently the quadrupole mode was excited. In other words, the trapped BEC was squeezed in one direction. The width of the BEC was then observed to exhibit oscillating motion, and its main axis of excitation was rotating with the period corresponding to the angular momentum of the condensate. For the photon fluid we expect a similar behavior but now cast in the language of synthetic magnetic fields. We start by taking as an input beam at $z=0$ an elliptic Laguerre-Gaussian beam with an angular momentum $\ell=1$ which is confined by a transversal symmetric and quadratic potential of the form $\Delta n= 0.0025 n_0 (k_0r)^2$. In Fig. \ref{chi3dyn} we show the dynamics of such a beam. The asymmetric excitation is seen to rotate as a function of $z$, which shows that the excitation is subject to a synthetic magnetic field. This rotation also shows that we have angular momentum in the beam as was noted by Chevy \textit{et al.} \cite{chevy_2000}. Note further that the results in Ref. \cite{kumar_2016} where the angular momentum of a ring-shaped Bose-Einstein condensate was found by measuring the precession of density perturbations can also be interpreted in the language of synthetic magnetic fields: The precession can be understood as an effect of a constant magnetic field. We expect the photon fluid to behave correspondingly, which is highlighted by comparing in Fig. \ref{chi3dyn} to the results of Kumar \textit{et al.} \cite{kumar_2016}.

\section{Summary and conclusions}\label{SumCon}

We have developed a theory of artificial gauge fields for photon fluids and the cases of both second-order and third-order optical nonlinearities. The resulting equations apply to weak excitations in the presence of pump fields carrying orbital angular momentum, and constitute a type of Bogoliubov theory. The resulting generally complex artificial gauge fields experienced by the weak excitations are an interesting generalization of previous cases and reflect the {\it PT}-symmetry properties of the underlying non-Hermitian Hamiltonian. We demonstrated numerically that the artificial gauge fields can produce synthetic magnetism, with illustrative examples of the observable consequences given for both second-order and third-order nonlinearities. 

Further exploration of these artificial gauge fields are planned for future publications. Key investigations include the generalization to nonlocal photon fluids and the physical effects arising from the possibility of complex valued flux tubes in, for example, parametric amplification involving beams with OAM. Although complex vector potentials did arise in the examples given in this paper, the resulting synthetic magnetic field was real.

\begin{acknowledgements}
We acknowledge valuable discussions with M. Valiente. D.F. acknowledges financial support from the European Research Council under the European Unions Seventh Framework Programme Grant No. (FP/2007-17172013)/ERC GA 306559 and EPSRC (U.K., Grant No. EP/J00443X/1). N.W acknowledges support from the EPSRC CM-CDT Grant No. EP/L015110/1. P.\"O acknowledges support from EPSRC (U.K., Grant No. EP/M024636/1).
\end{acknowledgements}

\end{document}